%% file: spatio-short.tex
\documentclass[12pt]{article}
\usepackage{epsfig}
\usepackage{amsfonts}
\begin{document}
\begin{titlepage}
\begin{center}

{\Large Spatio-temporal Chaos and Vacuum Fluctuations of Quantized Fields}

\vspace{1.cm} 

{\bf Christian Beck}

\vspace{1.cm}

School of Mathematical Sciences, Queen Mary,
University of London, Mile End Road, London E1 4NS.

\vspace{3cm}

\end{center}

\abstract{
We consider deterministic chaotic models of
vacuum fluctuations on a small (quantum gravity) scale.
As a suitable small-scale dynamics,
nonlinear versions of strings, so-called
`chaotic strings' are introduced. These can be used to
provide the `noise' for second quantization
of ordinary strings
via the Parisi-Wu approach of stochastic quantization.
Extensive numerical evidence is presented that
the vacuum energy of chaotic strings is minimized
for the numerical values of the observed
standard model parameters, i.e.
in this extended approach to second quantization concrete predictions for
vacuum expectations of dilaton-like fields and hence on masses
and coupling constants can be given. 
Low-energy fermion and boson masses are correctly obtained
with a precison of
3-4 digits, the electroweak and
strong coupling strengths with a precison of 4-5 digits. In
particular, the minima
of the vacuum energy  yield
high-precision predictions 
of the
Higgs mass ($154$ GeV), of the neutrino masses ($1.45\cdot10^{-5}$ eV,
$2.57\cdot 10^{-3}$ eV, $4.92 \cdot 10^{-2}$ eV)
and of the GUT scale ($1.73 \cdot 10^{16}$ GeV).}

\end{titlepage}

\newpage

The following text is preface, introduction, (detailed) summary, 
and
bibliography of the
book `Spatio-temporal Chaos
and Vacuum Fluctuations of Quantized Fields' by
C. Beck (World Scientific, 2002). All references to sections and chapters
refer to this book.

\input s-preface.tex

\newpage

\input s-intro.tex

\input s-chap12.tex

\input s-biblio.tex


\end{document}

%% file: s-preface.tex
\section*{Preface}

This book is written for an interdisciplinary readership of
graduate students and researchers interested in nonlinear
dynamics, stochastic processes, statistical mechanics on the one
hand and high energy physics, quantum field theory, string theory
on the other. In fact, one of the goals that I had in mind when
writing this book was to make particle physicists become
interested in nonlinear dynamics, and nonlinear physicists become
interested in particle physics. Why that? Didn't so far these two
subjects evolve quite independently from each other? So what is
this book about?

Mathematically, the subject of the book are coupled map lattices
exhibiting spatio-temporal chaotic behaviour. Physically, the
subject is a topic that lies at the heart of elementary particle
physics: There are about 25 free parameters in the standard model
of electroweak and strong interactions, namely the coupling
strengths of the three interactions, the fermion and boson masses,
and various mass mixing angles. These parameters are not fixed at
all by the standard model itself, they are just measured in
experiments, and a natural question is why these free parameters
take on the numerical values that we observe in nature and not
some other values. It will turn out that the answer is closely
related to certain distinguished types of coupled map lattices
that we will consider in this book as suitable models of vacuum
fluctuations. These dynamical systems, called `chaotic strings' in the
following, are observed to have minimum vacuum energy for the
observed standard model parameters. They
yield an extension of ordinary quantization schemes
which can account for the free parameters.

In this sense this book deals with both, nonlinear dynamics and
high energy physics. So far only very few original papers have
been published on this very new subject. With the current book I
hope to make these important new applications for coupled chaotic
dynamical systems accessible to a
broad readership.

The book consists of 12 chapters. The first few chapters will
mainly concentrate onto the theory of the relevant class of
coupled map lattices, their use for second quantization
purposes, and their physical interpretation in
terms of vacuum fluctuations.
In the later chapters concrete numerical results are
presented and these are then related to standard model
phenomenology. Sections marked with an asterisk can be omitted at
a first reading, these sections deal with interesting side issues
which, however, are not necessary for the logical development of
the following chapters. In view of the fact that (unfortunately!)
many readers may not have the time to read this book from the
beginning to the end, I included a very detailed summary as a
self-contained chapter 12. This summary contains the most
important concepts and results of this book and is written in a
self-consistent way, i.e. no knowledge of previous chapters is
required.

The research described in this book developed over a longer period
of time at various places. I started to work on the relevant types
of coupled map lattices during my stay at the Niels Bohr
Institute, Copenhagen, in 1992 and continued during a stay at the
University of Maryland in 1993. Some important numerical results,
now described in section 7.2 and 8.5, were obtained at the RWTH
Aachen in 1994 as well as during a visit to the Max Planck
Institute for Physics of Complex Systems, Dresden, in 1996. The
main part of the work was done at my home institute, the School of
Mathematical Sciences at Queen Mary, University of London, as well
as during long-term research visits to the Institute for
Theoretical Physics at the University of California at Santa
Barbara in 2000 and to the Newton Institute for Mathematical
Sciences at Cambridge in 2001. The hospitality that I enjoyed
during these visits was very pleasant, and the nice research
atmosphere was really inspiring.

The number of people from which I learned during the past years
and who thus indirectly contributed to this book is extremely
large--- too large to list all these individuals separately here!
So at this point let me just thank all of them in one go.


\begin{center}
                          London, February 2002

                          Christian Beck

\end{center}


%% file: s-intro.tex
\section*{Introduction}


This book deals with new applications for coupled map lattices in
quantum field theories and elementary particle physics. We will
introduce appropriate classes of coupled map lattices (so-called
`chaotic strings') as suitable spatio-temporal chaotic models of
vacuum fluctuations.

From a mathematical point of view, coupled map lattices are
high-di\-men\-sional nonlinear dynamical systems with discrete
space, discrete time and continuous state variables. They were for
the first time introduced by Kaneko in 1984 \cite{l1}. The
dynamics is generated by local maps that are situated at the sites
of a lattice. There can be various types of couplings between the
maps at the lattice sites, for example global coupling,
exponentially decreasing coupling or diffusive coupling. For
globally coupled systems, typically each lattice site is connected
to all others with the same coupling strength. In the
exponentially decreasing case the coupling strength decays
exponentially with distance. For diffusively coupled map lattices
there is just nearest-neighbor coupling, corresponding to a
discrete version of the Laplacian. The latter one is the most
relevant coupling form for applications in quantum field theories.
Very complicated periodic, quasi-periodic or spatio-temporal
chaotic behaviour is possible in all these cases (see the color
plates in chapter 2 and 4 for some illustrations).

Generally, the spectrum of possibilities of spatio-temporal
structures that can be generated by coupled map lattices is
extremely rich and has been extensively studied in the literature,
the emphasis being on the bifurcation structure
\cite{48,56,67}
\newline \cite{69,76,sym3}, Liapunov exponents \cite{44,24,91}
\newline \cite{85},
traveling waves \cite{16}
\newline \cite{32}, phase transition-like phenomena
\cite{78,34,37,47}
\newline \cite{51,71,74}
\newline \cite{huse, mackay2000}, the existence of
smooth invariant measures \cite{8,18,36}
\newline
\cite{59}, synchronization
\cite{1}
\newline \cite{3,4,12}
\newline \cite{14,25}, control \cite{9}
\newline \cite{10,15,29}
\newline \cite{49} and many other
properties. Applications for coupled map systems have been pointed
out for various subjects, among them hydrodynamic turbulence
\cite{physreve,97,hi,bottin}, chemical waves \cite{kapral},
financial markets \newline \cite{96}, biological systems
\cite{2}
\newline \cite{54,70,dens} and, at a much more fundamental level, for
quantum field theories \cite{94,98}. In this book we will
concentrate on the quantum field theoretical applications.

A possible way of embedding coupled map lattices into a general
quantum field theoretical context is via the Parisi-Wu approach of
stochastic quantization \cite{l5,l6,dam,gozzi,
namiki,bat,rumpf}
\newline \cite{ryang,breit,albeverio}. In this approach a
quantized field is described by a stochastic differential equation
evolving in a fictitious time coordinate. Essentially,
spatio-temporal Gaussian white noise is added to the classical
field equation in order to second quantize it. The fictitious time
is different from the physical time; it is an additional parameter
that is a useful tool for the quantization of classical fields.
Quantum mechanical expectations can be calculated as expectations
with respect to the realizations of the stochastic process. It is
now possible to generate the spatio-temporal Gaussian white noise
of the Parisi-Wu approach by a weakly coupled chaotic dynamics on
a very small scale. In particular, if we choose e.g.\ Tchebyscheff
maps to locally generate the `chaotic noise', the convergence to
Gaussian white noise under rescaling can be proved rigorously
\cite{bero,bill,chernov,888,beck3,chew,zygmund}. If we quantize by
means of such a chaotic dynamics, no difference occurs on large
(standard-model) scales, since on large scales the chaotic
behavior of the maps is very well approximated by Gaussian white
noise, leading to ordinary quantum field theoretical behavior.
However, on very small scales (e.g. the Planck scale or below)
there are interesting differences and new remarkable features. The
view that the ultimate theory underlying quantum mechanical
behaviour on a small scale is a deterministic one exhibiting
complex behaviour has also been advocated by `t Hooft
\cite{th2,thooft,thooft2}.

How can a discrete chaotic noise dynamics arise from an ordinary
field theory? How can there be a dynamical origin of the noise? We
will show that ordinary continuum field theories with formally
infinitely large self interaction directly and intrinsically lead
to diffusively coupled map lattices exhibiting spatio-temporal
chaos. This limit of large couplings stands in certain analogy to
the anti-integrable limit of Frenkel-Kontorova-like models
\cite{claude,aubry}. One of our main examples is a
self-interacting scalar field of $\phi^4$-type, which leads to
diffusively coupled cubic maps in the anti-integrable limit. A
discrete dynamics with strongest possible chaotic properties can
then be obtained, which can be used for stochastic quantization.
One can then consider coupled string-like objects in the noise
space, which, to have a name in the following, will be called
`chaotic strings'. We will use this model and some related ones as
dynamical models of vacuum fluc\-tua\-tions. The chaotic dynamics
will be scale invariant, similar as fully developed turbulent
states in hydrodynamics exhibit a selfsimilar dynamics on a large
range of scales \cite{turb1,turb4,procaccia,pope,ruelle}. In fact,
chaotic strings behave very much like a turbulent quantum state.
The probabilistic aspects of our model can be related to a
generalized version of statistical mechanics, the formalism of
nonextensive statistical mechanics
\cite{tsallis1,tsallis2,abe1,abe2,dyna,BLS,rad2,wilk,pennini,
johal,cohen2}.

What can we learn from these types of statistical models? We will
show that the assumption of a dynamical origin of vacuum
fluctuations, due to chaotic strings on a small scale, can help
to explain and reduce the large number of free parameters of the
standard model. The guiding principle for this is the minimization
of vacuum energy of the chaotic string. We will provide numerical
evidence that the vacuum energy is minimized for certain
distinguished string coupling constants. These couplings are
numerically observed to coincide with running standard model couplings
as well as with
gravitational couplings, taking for the energy scales the
masses of the known quarks, leptons, and gauge bosons. In this way
our approach can help to understand many of the free parameters of
the standard model, using concepts from generalized statistical
mechanics.

The approach described in this book is new and
different from previous attempts to calculate, e.g.,
the fine structure constant \cite{eddington,gil}.
It is much more in line with a
suggestion made by R.S. MacKay in his book \cite{bookrob} (p. 291), namely
that the fine structure constant
might be derived as a property of a fixed point of an appropriate
renormalization operator.
As we
shall see in chapter 7, the relevant dynamical systems are indeed the chaotic
strings, the renormalization operator
is a scale transformation,
and the renormalization flow corresponds to an evolution
equation for possible standard model couplings in the fictitious
time of the Parisi-Wu approach. This renormalization flow is not
only relevant for the
fine structure constant but provides information on all the other standard model
parameters as well.

The minima of the vacuum energy of chaotic strings can be
determined quite precisely and allow for high-precision
predictions of various running electroweak, strong, Yukawa and
gravitational coupling constants. These can then be translated
into high-precision estimates of the masses of the particles
involved. Moreover, evolving the couplings to higher energies
grand unification scenarios can be constructed. In this sense the
approach described in this book yields an interesting amendment of
the usual formulation of the standard model. Based on the
assumption that chaotic noise strings exist in addition to the
continuous standard model fields, we obtain high-precision
predictions of the free parameters of the standard model (see
Tab.~4 in chapter 12), which can be checked by experiments. Our
chaotic models yield rapidly evolving dynamical models of vacuum
fluctuations which, as we will show in detail in the following
chapters, have minimum vacuum energy for the observed standard
model parameters.

Can we further embed the chaotic strings into other theories, for
example superstring and M-theory
\cite{green,kaku,pol,pol1999,
wit,banks,
gauntlett}
\newline \cite{susskind,an1,witten3}, or relate them
to models of 2-dimensional quantum gravity \cite{gross} or string
cosmology \cite{ghosh,melch}
\newline \cite{ven1,lidsey}? Could the
very recently established contact between string field theory
and stochastic quantization yield a suitable embedding
\cite{polyakov,baulieu,periwal}
\newline \cite{ennyu}? All this is possible but
open at the moment.
Generally it
should be clear that chaotic strings are very different from
superstrings. The latter ones evolve in a regular way, the former
ones in a chaotic way. Still it is reasonable to look for possible
connections with candidate theories of quantum gravity, such as
superstring theory or M-theory. These theories require an
extension of ordinary 4-dimensional space-time to 10 (or 11)
space-time dimensions. The 6 extra dimensions are thought to be
`compactified', i.e. they are curled up on small circles with
periodic boundary conditions. One possible way to embed chaotic
strings is to assume that they live in the compactified space of
superstring theory. The couplings of the chaotic strings can then
be regarded as a kind of inverse metric in the compactified space,
determining the strength of the Laplacian coupling. The analogue
of the Einstein equations as well as suitable scalar field
equations then lead to the observed standard model coupling
constants, fixed and stabilized as equilibrium metrics in the
compactified space.

Let us give an overview over the following chapters. In chapter 1
we will generalize the stochastic quantization method to a chaotic
quantization method, where the noise is generated by a discrete
chaotic dynamics on a small time scale. In chapter 2 we will
introduce chaotic strings and discuss some of their symmetry
properties. Two types of vacuum energies associated with chaotic
strings are discussed in chapter 3, namely the self energy and the
interaction energy of chaotic strings. Spontaneous symmetry
breaking phenomena for chaotic strings and their
higher-dimensional extensions will be investigated in chapter 4.
In chapter 5 we will show why chaotic strings can be regarded as
simple selfsimilar dynamical models of vacuum fluctuations, and
introduce webs of Feynman graphs that describe this physical
interpretation. In chapter 6 we will relate the chaotic string
dynamics to a thermodynamic description of the vacuum, using
concepts from generalized statistical mechanics and information
theory. In chapter 7 we will consider analogues of Einstein field
equations that make {\em a priori} arbitrary standard model
couplings evolve to the stable zeros of the interaction energy of
chaotic strings. We will provide extensive numerical evidence that
the smallest stable zeros of the interaction energy numerically
coincide with running electroweak and strong coupling strengths,
evaluated at the smallest fermionic and bosonic mass scales. In
chapter 8 we will consider suitable self-interacting scalar field
equations for possible standard model couplings, which make {\em a
priori} arbitrary couplings evolve to the local minima of the self
energy of the chaotic strings. We will present extensive numerical
evidence that the self energy has local minima that numerically
coincide with various Yukawa, gravitational, electroweak and
strong couplings at energy scales given by masses of the three
families of quarks and leptons. In chapter 9 we extend the
analysis to bounded quark states, and provide numerical evidence
that the total vacuum energy has minima for running strong
coupling constants that correspond to the mass spectrum of light
mesons and baryons. The precision results of chapter 7 and 8 will
be used in chapter 10 to evolve the standard model couplings to
much higher energies and to construct grand unification scenarios.
In chapter 11 we will discuss the connection with extra dimensions
and describe possible scenarios at the Planck scale and beyond.
Finally, chapter 12 is a detailed, self-contained summary of the
most important concepts and results described in chapter 1-11.

%% file: s-chap12.tex
\section*{CHAPTER 12 SUMMARY}

\section*{12.1 Motivation and main results}

A fundamental problem of elementary particle physics is the fact
that there are about 25 free fundamental constants which are not
understood on a theoretical basis. These constants are essentially
the values of the three coupling constants, the quark and lepton
masses, the $W$ and Higgs boson mass, and various mass mixing
angles. An explanation of the observed numerical values is
ultimately expected to come from a larger theory that embeds the
standard model. Prime candidates for this are superstring and
M-theory \cite{green,kaku,pol,wit,banks}. However, so far the
predictive power of these and other theories is not large enough
to allow for precise numerical predictions.

In this book we have developed a dynamical theory of vacuum
fluctuations that may provide possible answers to this problem. We
have found that there is a simple class of 1+1-dimensional
strongly self-interacting discrete field theories (which, in order
to have a name, we have called `chaotic strings') that have a
remarkable property. The expectation of the vacuum energy of these
strings is minimized for string couplings that numerically
coincide with running standard model or gravitational couplings
$\alpha (E)$, the energy $E$ being given by the masses of the
known quarks, leptons, and gauge bosons. Chaotic strings can thus
be used to provide theoretical arguments why certain standard
model parameters are realized in nature, others are not. It is
natural to assume that the {\em a priori} free parameters evolve
to the local minima of the effective potentials generated by the
chaotic strings. Out of the many possible vacua, chaotic strings
seem to select the physically relevant states.

The dynamics of the chaotic strings is discrete in both space and
time and exhibits strongest possible chaotic behaviour \cite{98}.
It can be regarded as a dynamics of vacuum fluctuations that can
be used to 2nd-quantize other fields, for example ordinary
standard model fields, or ordinary strings, by dynamically
generating the noise of the Parisi-Wu approach of stochastic
quantization \cite{l5,damgaard} on a very small scale.
Mathematically, chaotic strings are 1-dimensional coupled map
lattices \cite{l1} of diffusively coupled Tchebyscheff maps $T_N$
of order $N$. The dynamics describes a kind of `turbulent quantum
state'. It turns out that there are six different relevant chaotic
string theories
---similar to the six components
that make up M-theory in the moduli space of superstring theory
\cite{gauntlett}. We have labeled these six chaotic string
theories as $2A,2B,2A^-,2B^-,3A,3B$. Here the first number denotes
the index $N$ of the Tchebyscheff polynomial and the letter A,B
distinguishes between a forward and backward coupling form. The
index $\,^-$ denotes anti-diffusive coupling (alternating signs of
Tchebyscheff polynomials in spatial direction). In principle one
can study these string theories for arbitrary $N$, but for
stochastic quantization only the cases $N=2$ and $N=3$ yield
non-trivial behaviour in a first and second order perturbative
approach \cite{hi2,hilnew}.


Chaotic strings can be used to generate effective potentials for
possible standard model couplings, regarding the {\em a priori}
free couplings as suitable scalar fields. The chaotic dynamics can
be embedded into ordinary physics in various ways, ranging from a
generalization of stochastic quantization (chapter 1) to an
extension of statistical mechanics (chapter 6) and to a quantum
gravity setting (chapter 11). Assuming that the {\em a priori}
free standard model couplings evolve to the minima of the
effective potentials generated by the chaotic strings, one can
obtain a large number of very precise predictions. The smallest
stable zeros of the expectation of the interaction energy of the
chaotic $3A$ and $3B$ strings are numerically observed to coincide
with the running electroweak couplings at the smallest fermionic
mass scales. Inverting the argument, the chaotic 3A string can be
used to theoretically predict that the low-energy limit of the
fine structure constant has the numerical value
$\alpha_{el}(0)=0.0072979(17)=1/137.03(3)$, to be compared with
the experimental value $1/137.036$. The $3B$ string predicts that
the effective electroweak mixing angle is numerically given by
$\bar{s}_l^2=sin^2\theta_{eff}^{lept}=0.23177(7)$, in perfect
agreement with the experimental measurements, which yield the
value $\bar{s}_l^2=0.23185(23)$ \cite{pada}. The smallest stable
zeros of the interaction energy of the $N=2$ strings are observed
to coincide with strong couplings at the smallest bosonic mass
scales. In particular, the smallest stable zero of the interaction
energy of the $2A$ string yields a very precise prediction of the
strong coupling at the $W$ mass scale, which, if evolved to the
$Z^0$ scale, corresponds to the prediction
$\alpha_s(m_{Z^0})=0.117804(12)$. The current experimentally
measured value is $\alpha_s(m_{Z^0})=0.1185(20)$ \cite{pada}.

Besides the coupling strengths of the three interactions, also the
fermion mass spectrum can be obtained with high precision from
chaotic strings. Here the expectation of the self energy of the
chaotic strings is the relevant observable. One observes a large
number of string couplings that locally minimize the self energy
and at the same time numerically coincide with various running
electroweak, strong, Yukawa and gravitational couplings, evaluated
at the mass scales of the higher fermion families. The highest
precision predictions for fermion masses comes from the self
energy of the $2A$ and $2B$ strings, which is observed to exhibit
minima for string couplings that coincide with gravitational and
Yukawa couplings of all known fermions. These minima of the vacuum
energy yield the free masses of the six quarks as $m_u=5.07(1)$
MeV, $m_d=9.35(1)$ MeV, $m_s=164.4(2)$ MeV, $m_c=1.259(4)$ GeV,
$m_b=4.22(2)$ GeV and $m_t=164.5(2)$ GeV. Note that a free top
mass prediction of 164.5(2) GeV corresponds to a top pole mass
prediction of 174.4(3) GeV, in very good agreement with the
experimentally measured value $M_t=174.3 \pm 5.1$ GeV. The masses
of the charged leptons come out as $m_e=0.5117(8)$ MeV,
$m_\mu=105.6(3)$ MeV and $m_\tau=1.782(7)$ GeV. All these
theoretically obtained values of fermion masses are in perfect
agreement with experimental measurements. To the best of our
knowledge, there is no other theoretical model that has achieved
theoretical predictions of similar precision. Chaotic strings also
provide evidence for massive neutrinos, and yield concrete
predictions for the masses of the neutrino mass eigenstates
$\nu_1,\nu_2,\nu_3$. These are $m_{\nu_1}=1.452(3)\cdot 10^{-5}$
eV, $m_{\nu_2}=2.574(3)\cdot 10^{-3}$ eV, $m_{\nu_3}=4.92(1)\cdot
10^{-2}$ eV (our symmetry considerations would also allow the mass
$m_{\nu_2}$ to be smaller by a factor $1/8$).

Not only fermion masses, but also boson masses can be obtained
from chaotic strings. The 2A string correctly reproduces the
masses of the $W$ and $Z$ boson, and a suitable interpretation of
the $2B^-$ string dynamics provides evidence for the existence of
a scalar particle of mass $m_H=154.4(5)$ GeV, which could be
identified with the Higgs particle. The latter mass prediction is
slightly larger than supersymmetric expectations but well within
the experimental bounds based on the ordinary standard model. We
also obtain estimates of the lightest glueball masses, which are
consistent with estimates from lattice QCD.

\section*{12.2 The chaotic string dynamics}



From a nonlinear dynamics point of view, chaotic strings are
easily introduced. They are 1-dimensional coupled map lattices of
diffusively coupled Tchebyscheff maps. In fact, for somebody with
a background in dynamical systems the dynamics is a
straightforward standard example of a spatially extended dynamical
system exhibiting chaotic behaviour. On the other hand, for
somebody working in high energy physics the equation may first
look somewhat unfamiliar, but the results summarized in the
previous section certainly indicate that it is worth learning new
things.

Consider a 1-dimensional lattice with lattice sites labeled by an
integer $i$. At each lattice site $i$ there is a variable
$\Phi_n^i$ that takes on values in the interval $[-1,1]$. $n$ is a
discrete time variable. Given some initial value $\Phi_0^i$ the
time evolution is given by deterministic recurrence relation
\begin{equation}
\Phi_{n+1}^i=T_N(\Phi_n^i).
\end{equation}
Here $T_N(\Phi)$ is the $N$-th order Tchebyscheff polynomial. One
has $T_2(\Phi)=2\Phi^2-1$ and $T_3(\Phi)=4\Phi^3-3\Phi$, generally
$T_N(\Phi )=\cos (N \arccos \Phi )$. The Tchebyscheff maps $T_N$
with $N\geq 2$ are known to exhibit strongly chaotic behaviour.
There is sensitive dependence on initial conditions: Small
perturbations in the initial values will lead to completely
different trajectories in the long-term run. The maps are
conjugated to a Bernoulli shift with an alphabet of $N$ symbols.
This means, in suitable coordinates the iteration process is just
like shifting symbols in a symbol sequence (see section 1.6 or any
textbook on dynamical systems for more details). As shown in
\cite{non1} Tchebyscheff maps have least higher-order correlations
among all systems conjugated to a Bernoulli shift, and are in that
sense closest to Gaussian white noise, though being completely
deterministic. A graph theoretical method for this type of
`deterministic noise' has been developed in \cite{non1, hilnew}.

We now couple the Tchebyscheff dynamics with a small coupling $a$
in the spatial direction labeled by $i$, obtaining the chaotic
string dynamics:
\begin{equation}
\Phi_{n+1}^i=(1-a)T_N(\Phi_n^i) + s \frac{a}{2} (T_N^b
(\Phi_n^{i-1}) +T_N^b(\Phi_n^{i+1})) \label{sum1}
\end{equation}
We consider both the positive and negative Tchebyscheff polynomial
$T_{\pm N} (\Phi)$ $=\pm T_N(\Phi)$, but have suppressed the index
$\pm$ in the above equation. The variable $a$ is a coupling
constant taking values in the interval $[0,1]$. $s$ is a sign
variable taking on the values $\pm1$. The choice $s =+ 1$ is
called `diffusive coupling', but for symmetry reasons it also
makes sense sense to study the choice $s=-1$, which we call
`anti-diffusive coupling'. The integer $b$ distinguishes between
the forward and backward coupling form, $b=1$ corresponds to
forward coupling ($T_N^1(\Phi):=T_N(\Phi))$, $b=0$ to backward
coupling ($T_N^0(\Phi):=\Phi$). We consider periodic boundary
conditions and large lattices of size $i_{max}$.

The dynamics (\ref{sum1}) is deterministic chaotic, spatially extended,
and strongly nonlinear. The field variable $\Phi_n^i$
is physically interpreted in terms of rapidly fluctuating
virtual momenta in units of some arbitrary maximum
momentum scale. There are some analogies with velocity fluctuations
in fully developed turbulent flows, which are also
deterministic chaotic, spatially extended, and induced by strong
nonlinearities. For this reason it makes sense to think
of eq.~(\ref{sum1}) as describing a turbulent state of
vacuum fluctuations, or in short a `turbulent quantum state'.
These states appear to have physical relevance, since they
reproduce the observed values of the standard model parameters.

We may also write the coupled dynamics as
\begin{equation}
\Phi_{n+1}^i=T_N(\Phi_n^i)+\frac{a}{2} \left(
sT_N^b(\Phi_n^{i-1})-2T_N(\Phi_n^i)+sT_N^b(\Phi_n^{i+1}) \right) .
\end{equation}
This way of writing illustrates that the effect of the coupling is
similar to the action a Laplacian operator. Since $a$ determines
the strength of the Laplacian coupling, and since in quantum field
theories this role is usually attributed to a metric,  $a^{-1}$
can be regarded as a kind of metric in the 1-dimensional string
space indexed by $i$.

It is easy to see that for odd $N$ the statistical properties of
the coupled map lattice are independent of the choice of $s$
(since odd Tchebyscheff maps satisfy $T_N(-\Phi )=-T_N(\Phi )$),
whereas for even $N$ the sign of $s$ is relevant and a different
dynamics arises if $s$ is replaced by $-s$. Hence, restricting
ourselves to $N=2$ and $N=3$, in total 6 different chaotic string
theories arise, characterized by
$(N,b,s)=(2,1,+1),(2,0,+1),(2,1,-1),(2,0,-1)$ and
$(N,b)=(3,1),(3,0)$. For easier notation, we have labeled these
string theories as $2A,2B,2A^-,2B^-,3A,3B$, respectively.

If the coupling $a$ is sufficiently small, the chaotic variables
$\Phi_n^i$ can be used to generate the noise of the Parisi-Wu
approach of stochastic quantization \cite{l5,l6} on a very small
scale. That is to say, we assume that on a very small (quantum
gravity) scale the noise used for quantization purposes is not
structureless but evolves in a deterministic chaotic way. Just on
a large scale it looks like Gaussian white noise, and there is
convergence to ordinary path integrals using this `deterministic
noise', as can be rigorously proved for $a = 0$ \cite{bero,
beck3,beck44}. In this interpretation the discrete time variable
$n$ corresponds to the fictitious time of the Parisi-Wu approach,
an artificial time coordinate that is just introduced for
quantization purposes (see chapter 1 for details).


The chaotic string dynamics (\ref{sum1}) formally originates from
a 1-dimen\-sional continuum $\phi^{N+1}$-theory in the limit of
infinite self-interaction strength (see section 2.2 for details).
In this sense, chaotic strings can also be regarded as degenerated
Higgs-like fields with infinite self-interaction parameters, which
are constraint to a 1-dimensional space. Another way to physically
motivate chaotic strings is to emphasize certain analogies with
ordinary strings (section 2.4), to connect them with fluctuating
momenta that are allowed due to the uncertainty relation (sections
5.1-5.3), and to relate them to a 1+1 dimensional model of quantum
gravity (section 5.7). One can also embed them into the
compactified space of string theories (section 11.1).

\section*{12.3 Vacuum energy of chaotic strings}

Though the chaotic string dynamics is dissipative, one can
formally introduce potentials that generate the discrete time
evolution. For $a=0$ we may write
\begin{equation}
\Phi_{n+1}-\Phi_n=\pm T_N(\Phi_n)=-\frac{\partial}{\partial \Phi_n}V_\pm
(\Phi_n).
\end{equation}
For $N=2$ the (formal) potential is given by
\begin{equation}
V^{(2)}_\pm (\Phi )= \pm \left( -\frac{2}{3} \Phi^3 +\Phi \right)
+\frac{1}{2} \Phi^2+C, \label{sumv2}
\end{equation}
for $N=3$ by
\begin{equation}
V^{(3)}_\pm (\Phi )=\pm \left( -\Phi^4+\frac{3}{2} \Phi^2 \right)
+\frac{1}{2} \Phi^2 +C . \label{sumv3}
\end{equation}
Here $C$ is an arbitrary constant. The uncoupled case $a=0$ is
completely understood. The dynamics is ergodic and mixing. Any
expectation of an observable $A(\Phi)$ can be calculated as
\begin{equation}
\langle A \rangle = \int_{-1}^1 d\phi \; \rho (\phi)  A(\phi ),
\end{equation}
where
\begin{equation}
\rho (\phi )=\frac{1}{\pi \sqrt{1-\phi^2}}  \label{sumberlin}
\end{equation}
is the natural invariant density describing the probability
distribution of iterates of Tchebyscheff maps (see, e.g.,
\cite{bes} for an introduction). In more general versions of
statistical mechanics \cite{tsallis1}, this probability density
can be regarded as a generalized canonical distribution (see
section 6.4 for more details).

If a spatial coupling $a$ is introduced, things become much more
complicated, and the invariant 1-point density deviates from the
simple form (\ref{sumberlin}). A spatial coupling is formally
generated by the interaction potential $aW_\pm (\Phi , \Psi )$,
with
\begin{equation}
W_\pm (\Phi , \Psi ) = \frac{1}{4} (\Phi \pm \Psi )^2 +C.
\label{sumw}
\end{equation}
Here $\Phi$ and $\Psi$ are neighbored noise field variables on the
lattice. One has
\begin{equation}
-\frac{\partial}{\partial \Phi^i}W_\pm (\Phi^i , \Phi^{i+1})
-\frac{\partial}{\partial \Phi^i}W_\pm (\Phi^i , \Phi^{i-1})
= \pm \frac{1}{2} \Phi^{i+1}-\Phi^i \pm \frac{1}{2} \Phi^{i-1}.
\end{equation}
This generates diffusive $(+)$, respectively anti-diffusive $(-)$
coupling. Anti-diffusive coupling can equivalently be obtained by
keeping $W_-$ but replacing $T_N\to -T_N$ at odd lattice sites.
The coupled map dynamics (\ref{sum1}) is obtained by letting the
action of $V$ and $W$ alternate in discrete time $n$, then
regarding the two time steps as one.

The expectations of the potentials $V$ and $W$ yield two types of
vacuum energies $V_\pm (a):=\langle V_\pm^{(N)}(\Phi^i ) \rangle$ (the
self energy) and $W_\pm (a):=\langle W_\pm (\Phi^i ,\Phi^{i+1} )
\rangle$ (the interaction energy). Here $\langle \cdots \rangle$
denotes the expectation with respect to the coupled chaotic
dynamics. Numerically, any such expectation can be determined by
averaging over all $i$ and $n$ for random initial conditions
$\Phi^i_0\in [-1,1]$, omitting the first few transients. Note that
in the stochastic quantization approach the chaotic
noise is used for 2nd quantization of standard model fields (or
ordinary strings) via the Parisi-Wu approach \cite{98}. Hence generally
expectations with respect to the chaotic dynamics correspond to
expectations with respect to 2nd quantization. The expectation of
the vacuum energy of the string, given by the above functions
$W_\pm (a)$ and $V_\pm (a)$, depends on the coupling $a$ in a non-trivial
way. Moreover, it also depends on the integers $N,b,s$ that define
the chaotic string theory.

Since negative and positive Tchebyscheff maps essentially generate
the same dynamics, up to a sign, any physically relevant
observable should be invariant under the transformation $T_N\to
-T_N$. The vacuum energies $V_\pm (a)$ and $W_\pm (a)$ of the various
strings exhibit full symmetry
under the transformation $T_N \to -T_N$ (respectively $s \to -s$) if the
additive constant $C$ is chosen to be
\begin{equation}
C= -\frac{1}{2} \langle \Phi^2 \rangle.
\end{equation}
For that choice of $C$, the expectations of $V_+$ and $V_-$ as well
as those of $W_-$ and $W_+$ are the same, up
to a sign.

Choosing (by convention) the $+$ sign one obtains from
eq.~(\ref{sumv2}), (\ref{sumv3}) and (\ref{sumw}) for the
expectations of the potentials
\begin{eqnarray}
V^{(2)}(a) &=& -\frac{2}{3} \langle \Phi^3 \rangle +\langle \Phi
\rangle \\ V^{(3)}(a) &=&  -\langle \Phi^4 \rangle +\frac{3}{2}
\langle \Phi^2 \rangle ,
\end{eqnarray}
and
\begin{equation}
W(a) =\frac{1}{2} \langle \Phi^{i} \Phi^{i+1} \rangle .
\end{equation}

The above $\pm$ symmetry can actually be used to cancel unwanted
vacuum energy and to avoid problems with the cosmological
constant. If one assumes that strings with both $T_N$ and $-T_N$
are physically relevant, the two contributions $V(a)$ and $-V(a)$
(respectively $W(a)$ and $-W(a)$) may simply add up to zero. This
reminds us of similarly good effects that supersymmetric partners
have in ordinary quantum field and string theories. In fact, as a
working hypothesis we may regard the above $Z_2$ symmetry as
representing a kind of supersymmetry in the chaotic noise space.

Similarly as for ordinary strings it also makes sense to consider
certain conditions of constraints for the chaotic string. For
ordinary strings (for example bosonic strings in covariant gauge
\cite{green}) one has the condition of constraint that the energy
momentum tensor should vanish. The first diagonal component of the
energy momentum tensor is an energy density. For chaotic strings,
the evolution in space $i$ is governed by the potential $W_\pm
(\Phi, \Psi)$ and the corresponding expectation of the energy
density is $\pm W(a)$. We should thus impose the condition of
constraint that $W(a)$ should vanish for physically observable
states. Moreover, the evolution in fictitious time $n$ is governed
by the self-interacting potential $V^{(N)} (\Phi )$. This
potential generates a shift of information, since the Tchebyscheff
maps $T_N$ are conjugated to a Bernoulli shift of $N$ symbols.
Hence $V(a)$ can be regarded as the expectation of a kind of
information potential or entropy function, which, motivated by
thermodynamics, should be extremized for physically observable
states. Note that the action of $V$ and $W$ alternates in $n$ and
$i$ direction. Both types of vacuum energies describe different
relevant observables of the chaotic string and are of equal
importance.

\section*{12.4 Fixing standard model parameters}

In order to
construct a link to standard model phenomenology it is useful to
introduce a simple physical interpretation of the chaotic string
dynamics in terms of fluctuating virtual momenta.
Suppose we regard $\Phi_n^i$ to be a fluctuating virtual momentum
component that can be associated with a hypothetical particle $i$
at time $n$ that lives in the constraint 1-dimensional string
space. $n$ can be either interpreted as fictitious time or as
physical time, it doesn't really matter for our purposes.
Neighbored particles $i$ and $i-1$ can exchange momenta due to the
Laplacian coupling of the coupled map lattice. To make this model
more concrete we may assume that at each time step $n$ a
fermion-antifermion pair $f_1,\bar{f}_2$ is spontaneously created
in cell $i$. In units of some arbitrary energy scale $p_{max}$,
the particle has momentum $\Phi_n^i$, the antiparticle momentum
$-\Phi_n^i$. They interact with particles in neighbored cells by
exchange of a (hypothetical) gauge boson $B_2$, then they
annihilate into another boson $B_1$ until the next vacuum
fluctuation takes place. This can be symbolically described by the
Feynman graph in Fig.~1.
\begin{figure}
\epsfxsize=1. \hsize
\epsfig{file=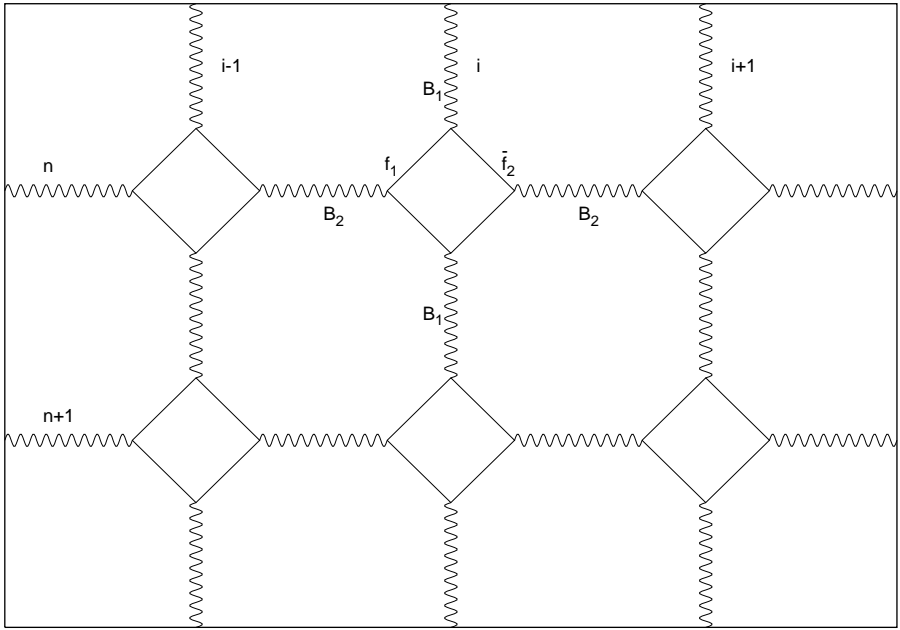}
\caption{Interpretation of the coupled map dynamics in terms of
fluctuating momenta exchanged by fermions $f_1,\bar{f}_2$ and
bosons $B_1,B_2$.}
\end{figure}
We called this graph a `Feynman web', since it describes an
extended spatio-temporal interaction state of the string, to which
we have given a standard model-like interpretation. Note that in
this interpretation $a$ is a (hypothetical) standard model
coupling constant, since it describes the strength of momentum
exchange of neighbored particles. At the same time, $a$ can also
be regarded as an inverse metric in the 1-dimensional string
space, since it determines the strength of the Laplacian coupling.

What is now observed numerically for the various chaotic strings
is that the interaction energy $W(a)$ has zeros and the self energy $V(a)$
has local minima for string couplings $a$ that numerically coincide
with running standard model couplings
$\alpha(E)$, the energy
being given by
\begin{equation}
E= \frac{1}{2} N \cdot (m_{B_1}+m_{f_1}+m_{f_2}). \label{sumka}
\end{equation}
Here $N$ is the index of the chaotic string theory considered, and
$m_{B_1}, m_{f_1}$, $m_{f_2}$ denote the masses of the standard
model particles involved in the Feynman web interpretation. The
surprising observation is that rather than yielding just some
unknown exotic physics, the chaotic string spectrum appears to
reproduce the masses and coupling constants of the known quarks,
leptons and gauge bosons of the standard model with very high
precision. Gravitational and Yukawa couplings are observed as
well.

We thus have the possibility to fix and predict the free standard
model parameters by simply assuming that the entire set of
parameters is chosen in such a way that it minimizes the vacuum
energy of the chaotic strings.

Although eq.~(\ref{sumka}) looks like a low-energy formula, the
chaotic strings may well describe a scenario that takes place at
very large energies well above the Planck scale. The reason is
that alternatively the Feynman web dynamics can also be
interpreted in terms of black holes that emit low-energy particles
by Hawking radiation, similar to the process sketched in
Fig.~2.
\begin{figure}
\epsfxsize=1. \hsize 
\epsfig{file=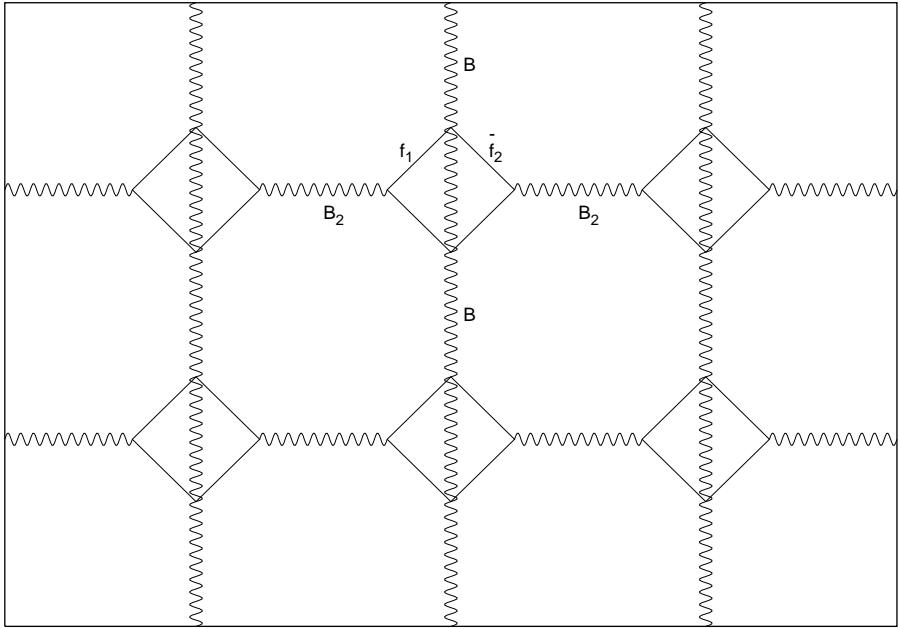} \caption{Alternative
interpretation of the coupled map dynamics. Black holes
(denoted by $B$) emit low-energy particles $f_1,\bar{f}_2$ that
interact with gauge bosons $B_2$ at low temperatures.}
\end{figure}
In this case a very large energy is given by something of the
order of the mass $M$ of the black holes, but what is physically
relevant is the Hawking temperature $kT_H\sim  \frac{1}{GM}$ by
which the black holes radiate. Here $G$ is the gravitational
constant. Thus very large energies $M$ can be associated with very
small Hawking temperatures $kT_H$ and allow for fixing of
low-energy parameters in a pre-Planck epoch (see section 11.5 for
more details).


At a very early stage of the universe, where standard model
parameters are not yet fixed and ordinary space-time may
not yet exist as well, pre-standard model couplings may be realized as
coupling constants $a$ in the chaotic string space. The parameters
are then fixed by an evolution equation (a renormalization flow)
of the form
\begin{equation}
\dot{a} = const \cdot W(a) + noise, \label{sum101}
\end{equation}
respectively
\begin{equation}
\dot{a}=-const \cdot \frac{\partial V}{\partial a} + noise ,
\label{sum102}
\end{equation}
where we assume that the constant $const$ is positive. It can also
depend on $a$, this doesn't matter as long as it does not switch
sign. The equations make the expectations of {\em a priori}
arbitrary standard model couplings $a$ evolve to the stable zeros
of $W(a)$, respectively to the local the minima of $V(a)$.
Eq.~(\ref{sum101}) is a kind of Einstein equation and
eq.~(\ref{sum102}) a kind of scalar self-interacting field
equation for $a$ (see sections 7.1, 8.1, 11.2 for more details).

\section*{12.5 Numerical findings}

Our numerical results for zeros of $W(a)$
and local minima of $V(a)$
and the corresponding physical interpretations are
described in detail in chapter 7-10. Let us here just summarize
the most important points.

The smallest non-trivial zeros of the interaction energy $W(a)$
with negative slope at the zero, describing stable stationary
states of couplings under the evolution equation~(\ref{sum101}),
are listed in Tab.~1.
\begin{table}

\begin{tabular}{|l|l|l|} \hline
string & stable zero & running SM coupling
\\ \hline 3A & $a_1^{(3A)}=0.0008164(8)$ &
$\alpha_{el}^d(3m_d)=0.0008166$ \\ 3A & $a_2^{(3A)}=0.0073038(17)$ &
$\alpha_{el}^e(3m_e)=0.007303$  \\ 3B
 & $a_1^{(3B)}=0.0018012(4)$ &
 $\alpha_{weak}^{u_R}(3m_u)+\alpha_{el}^d(3m_d)=0.001800$
 \\

3B & $a_2^{(3B)}=0.017550(1)$ &
$\alpha_{weak}^{\nu_L}(3m_{\nu_e})+\alpha_{el}^e(3m_e)=0.01755$ \\
2A &$a_1^{(2A)}=0.120093(3)$ &$\alpha_s (m_W+2m_d)=0.1208(20)$ \\
2B &$a_1^{(2B)}=0.3145(1)$ &$\alpha_s(m_{gg^{0++}}+2m_u)=?$ \\
$2A^-$ & $a_1^{(2A^-)}=0.1758(1)$ &
$\alpha_s(m_{gg^{2++}}+2m_b)$=? \\ $2B^-$ &
$a_1^{(2B^-)}=0.095370(1)$ & $\alpha_s(m_H+2m_t)$=? \\ \hline

\end{tabular}

\vspace{0.5cm}

\caption{The smallest stable zeros of the interaction energies of
the $3A,3B,2A,2B,2A^-,2B^-$ strings and comparison with
experimentally measured standard model couplings (where known).}

\end{table}
The $N=3$ zeros are observed to numerically coincide with the
coupling constants of electroweak interactions, evaluated at four
different energy scales. The four relevant energy scales are given
by the masses of the lightest fermions $d,e,u,\nu_e$. To calculate
the concrete value of the energy scale, one uses
eq.~(\ref{sumka}). For example, the zero $a_2^{(3A)}$ is
associated with a Feynman web of the form $f_1=e^-,\bar{f}_2=e^+$,
$B_1$ massless, $B_2= \gamma$. This yields
$E=\frac{3}{2}(2m_e+m_{B_1})=3m_e$. Tab.~1 shows that there is
coincidence of all four observed stable zeros with the
corresponding standard model couplings with a precision of 4
digits (more details in section 7.2--7.4).

Symmetry considerations with the $N=3$ strings suggest that the
smallest stable zeros of the $N=2$ strings fix strong couplings at
the smallest bosonic mass scales. These are given by the $W$
boson, the Higgs boson, and the lightest glueballs of spin 0 and
2. Since only the $W$ boson mass is precisely known, we can only
compare the zero $a_1^{(2A)}$ with experimental data. It indeed
coincides with $\alpha_s(m_W)$. The other zeros yield predictions
for the Higgs and glueball masses (details in section 7.6 and
7.7).

The observed zeros of the $N=2$ and $N=3$ strings allow for
high-precision predictions of the fine structure constant, the
Weinberg angle and the strong coupling constant. The fine
structure constant and the effective Weinberg angle for
$Z^0$-lepton coupling are correctly obtained with a precision of
4-5 digits. The strong coupling constant at the $W$ mass scale is
predicted with about 5 valid digits, a much higher precision
than can be confirmed by experiments at the present stage. $\alpha_s(E)$
can then be easily evaluated at other energy scales as well, using the
well-known QCD formulas.

Generally, we observe interesting symmetries between the various
zeros and the corresponding standard model interactions. These
kind of symmetries often help to fill `gaps' in the table, i.e.\
to find the right Feynman web interpretation for a given zero. For
the $N=3$ strings, proceeding from the smallest stable zeros
$a_1^{(3A/3B)}$ to the next larger stable zeros $a_2^{(3A/3B)}$
means replacing quarks by leptons, whereas for the $N=2$ string it
means replacing ordinary gauge and Higgs bosons ($W^\pm$ and $H$)
by their Planck-scale analogues ($\tilde{W}^\pm$ and $\tilde{H}$,
see section 10.6). For the $N=3$ strings, the forward coupling
form ($3A$) describes just one species of fermions (either $d$ or
$e$). These fermions certainly have spin. On the other hand, the
backward coupling form ($3B$) describes simultaneous states of two
fermions ($(u_R,d_L)$ and $(\nu_L ,e_R)$) of opposite handedness,
which coexist independently. Hence proceeding from forward to
backward coupling means proceeding from a state with spin to a
state with total spin $0$, since $R+L=0$. For the $N=2$ strings
proceeding from forward to backward coupling also means going from
a state with spin (the $W$-boson, the glueball with spin 2) to a
state without spin (the Higgs boson, the glueball with spin 0).



Now let us look at the other relevant type of vacuum energy, the
self energy $V(a)$. The functions $V(a)$ have plenty of local
minima for the various strings. For small $a$, oscillating scaling
behaviour sets in and all local minima of $V(a)$ are only fixed
modulo $N^2$ (details in section 8.6). Also, in this limit there
is no visible difference between the self energies of the forward
and backward coupling form. The self energy of the $2A/B$ string
turns out to have local minima for Yukawa couplings of heavy
fermions ($t,b,\tau,c$) and gravitational couplings of light
fermions ($\mu , s, d, u, e$). This is shown in Tab.~2.
The Yukawa coupling of the $t$ quark falls out of the scaling
region and is separately listed in Tab.~3.
\begin{table}

\begin{tabular}{|l|c|l|} \hline
local minimum & fermion & SM gravitational/Yukawa coupling
\\ \hline
$b_1=0.000199(1)$ mod 4 & $\mu$ & $\alpha_{G}^\mu =
0.0001989$ mod 4  \\
 $b_2=0.000263(2)$ mod 4 &$\tau$&$\alpha_{Yu}^\tau
=0.0002615$ mod 4 \\
 $b_4=0.000306(1)$ mod 4 & $e$ &
 $\alpha_{G}^e
=0.0003052$ mod 4\\   $b_6=0.000368(3)$ mod 4 & $b$&
$\alpha_{Yu}^b =0.000369(6)$ mod 4
\\  $b_7=0.000399(1)$ mod 4 & $d$& $\alpha_G^d=
0.00038(5)$ mod 4 \\  $b_8=0.000469(1)$ mod
4 & $u$& $\alpha_G^u=0.00040(17)$ mod 4
\\  $b_9=0.000482(1)$ mod 4 & $s$& $\alpha_G^s=
0.00050(4)$ mod 4 \\  $b_{10}=0.000525(2)$
mod 4 & $c$& $\alpha_{Yu}^c=0.000526(25)$
mod 4
\\ \hline

\end{tabular}

\vspace{0.5cm}

\caption{Local minima of the self energy of the 2A/B string in the
scaling region and comparison with gravitational and Yukawa
couplings of standard model particles. For the experimental quark
mass values that lead to the numbers and error bars in column 3 we
have chosen $m_t=164(5)$ GeV, $m_b=4.22(4)$ GeV, $m_c=1.26(3)$
GeV, $m_s=167(7)$ MeV, $m_d=9.1(6)$ MeV, $m_u=4.7(9)$ MeV.}

\end{table}

Clearly, since the minima of the vacuum energy can be determined
with high precision, and since gravitational and Yukawa couplings
depend quadratically on the mass, the minima allow for
high-precision predictions of masses modulo 2, in particular of
quark masses. Note that most of the values listed in column 1 of
Tab.~2 have a much higher precision than the experimental
values in column 3. There are further minima (not listed in
Tab.~2) that can be used to make neutrino mass predictions (see
section 8.7). For the attribution of the various minima to the
various particles one uses simple discrete symmetry
considerations.

For all states in the scaling region, the relevant power of 4 of
the coupling (equivalent to a power of 2 for the mass) is {\em a
priori} undetermined. It can be theoretically related to a folding
number of the string in a quantum gravity epoch (section 11.6). In
this epoch all masses are only fixed modulo 2. For standard model
states, however, the relevant power of 2 can be deduced by
postulating compatibility with other string states outside the
scaling region (see section 8.6--8.7 for details).


Outside the scaling region, lots of other minima are observed that
can directly be identified with standard model interaction
strengths (Tab.~3).
\begin{table}
\begin{tabular}{|l|l|l|} \hline
string & local minimum & running SM coupling
\\ \hline 3A & $a_1'^{(3A)}=0.000246(2)$ &
$\alpha_{weak}^{d_R}(3m_d)=0.000246$ \\ 3A &
$a_2'^{(3A)}=0.00102(1)$  & $\alpha_{weak}^{c_R}(3m_c)=0.000101$
\\ 3A
 & $a_3'^{(3A)}=0.00220(1)$ &$\alpha_{weak}^{e_R}(3m_e)=0.00220$ \\
3A & $a_6'^{(3A)}=0.0953(1)$  & $\alpha_s(3m_t)=0.0952(3)$  \\ 3A
& $a_7'^{(3A)}=0.1677(5)$ & $\alpha_s(3m_b)=0.1684(4)$  \\ 3A &
$\alpha_8'^{(3A)}=0.2327(5)$ &$\alpha_s(3m_c)=0.232(2)$  \\ 3B & $
\alpha_6'^{(3B)}=0.1027(1)$ &
$\alpha_s\left(\frac{3}{2}(m_t+m_b)\right) =0.1027(4)$ \\ 3B & $
\alpha_8'^{(3B)}=0.2916(5)$ &
$\alpha_s\left(\frac{3}{2}(m_c+m_s)\right) =0.291(4)$ \\ 2A &
$a_2'^{(2A)}=0.03369(2)$ & $\alpha_2(m_Z+2m_b)=0.03369(1)$ \\ 2B &
$a_2'^{(2B)}=0.03440(2)$ & $\alpha_{Yu}^t(m_H+2m_t)=0.0342(21)$ \\
\hline

\end{tabular}

\vspace{0.5cm}

\caption{Various observed local minima of the self energy of the
$3A,3B,2A,2B$ strings outside the scaling region and comparison
with running standard model couplings. The running strong coupling
$\alpha_s(E)$ can be very precisely evaluated using the zero
$a_1^{(2A)}$ listed in Tab.~1.}

\end{table}
For the $N=3$ strings we observe minima describing strongly
interacting heavy quarks $q$, i.e.\ $f_1=q$, $\bar{f}_2=\bar{q}$,
$B_1$ massless, $B_2=g$ (gluon), where $q=t,b,c$, respectively.
Pure flavor states are described by forward coupling and mixed
flavor states $(t,b)$ and $(c,s)$ by backward coupling. Further
minima can be identified with weak interaction states of
right-handed fermions, fixing the three different charges of
charged quarks and leptons. Small differences between the forward
and backward coupling form may possibly provide information on
mass mixing angles, i.e.\ the entries of the Kobayashi-Maskawa
matrix.
Between small (weak) and large (strong) couplings we have
identified the remaining two minima as describing unified
couplings at the GUT and Planck scale (see chapter 10).

Summarizing, the two types of vacuum energy and the six types of
chaotic strings considered seem to be sufficient to fix the most
relevant parameters of the standard model such as masses, charges
and coupling constants. The standard model appears to have evolved
to a state of minimum vacuum energy with respect to the chaotic
strings. Tab.~4 summarizes all predictions on standard model
parameters (together with the relevant error bars) that we have
extracted from the zeros and minima of the vacuum energy. For
comparison, also the experimentally measured values are listed.
\begin{table}

\begin{tabular}{|l|l|l|}
\hline parameter & chaotic string prediction & measured \\ \hline
$\alpha_{el}(0)$ &0.0072979(17) & 0.00729735253(3)=1/137.036 \\
$\bar{s}_l^2$ & 0.23177(7) &0.2318(2) \\ $\alpha_s(m_Z)$ &
0.117804(12) & 0.1185(20) \\ $m_e$ & 0.5117(8) MeV & 0.51099890(2)
MeV \\ $m_\mu$ & 105.6(3) MeV & 105.658357(5) MeV \\ $m_\tau$ &
1.782(7) GeV & 1.7770(3) GeV\\ $m_{\nu_1}$ & $1.452(3) \cdot
10^{-5}$ eV  & ? \\ $m_{\nu_2}$ & $2.574(3) \cdot 10^{-3}$ eV & ?
\\ $m_{\nu_3}$ & $4.92(1) \cdot 10^{-2}$ eV & $\sim 5 \cdot
10^{-2}$ eV ? \\ $m_u$ & 5.07(1) MeV & $\sim 5$ MeV
\\ $m_d$ & 9.35(1) MeV & $\sim 9$ MeV \\ $m_s$ & 164.4(2) MeV &
$\sim 170$ MeV \\ $m_c$ & 1.259(4) GeV & 1.26(3) GeV\\ $m_b$ &
4.22(2) GeV & 4.22(4) GeV\\ $m_t$ & 164.5(2) GeV & 164(5) GeV\\
$m_W$ & 80.36(2) GeV & 80.37(5) GeV\\ $m_H$ & 154.4(5) GeV & ? \\
\hline
\end{tabular}

\vspace{0.5cm}

\caption{Most important standard model parameters as obtained from
the chaotic string spectrum and as measured in experiments. All
quark masses denote free quark masses. $m_t=164.5(2)$ GeV
corresponds to a top pole mass of $M_t=174.4(3)$ GeV. Symmetry
considerations also allow 
$m_{\nu_2}$ to be smaller by a factor $\frac{1}{8}$ (see
section 8.7).}

\end{table}

Some predictions made by the chaotic strings are much more precise
than current experiments can confirm (e.g.\ $\alpha_s(m_W)$, the
free quark masses, neutrino masses, the Higgs mass), whereas other
predictions are less precise than the experimental data (e.g.\
$\alpha_{el}(0)$ and the charged lepton masses). All known masses
are correctly reproduced with a precision of 3-4 digits. The
unknown masses (Higgs and neutrino masses) are predicted with a
similar precision, based on simple symmetry arguments on what type
of minimum is relevant (see sections~7.7 and 8.7 for details).

One can also study the total vacuum energy of the strings as given
by $H_-(a)=V(a)-aW(a)$ and $H_+(a)=V(a)+aW(a)$. It turns out that
for the $N=3$ strings local minima of $H_-(a)$ essentially
reproduce the spectrum of light mesons (Tab.~5) and local
minima of $H_+(a)$ essentially that of light baryons (Tab.~6).
\begin{table}

\begin{tabular}{|l|c|l|} \hline
string & local minimum of $H_-(a)$ & strong coupling
\\ \hline 3A & 0.163 &
$\alpha_s(\frac{3}{2}m_{\Upsilon (1s)})=0.164$ \\ 3A & 0.186 &
$\alpha_s(\frac{3}{2}m_B)=0.188$   \\ 3A
 & 0.282 &$\alpha_s(\frac{9}{4}m_\Phi)=0.282$  \\
3A & 0.375 & $\alpha_s(\frac{3}{2}m_{K^*})=0.374$ \\ 3A & 0.416 &
$\alpha_s(\frac{3}{2}m_\rho )=0.418$
\\ 3A & 0.615 &
$\alpha_s(\frac{3}{2}m_\eta )=0.60$ \\ 3A & 0.676
&$\alpha_s(\frac{3}{2}m_{K} )=0.70$  \\ 3B & 0.161 &
$\alpha_s(\frac{3}{2}m_{\Upsilon (2s)})=0.161$ \\ 3B & 0.182 &
$\alpha_s(\frac{3}{2}m_{B_c})=0.179$ \\ 3B & 0.224 &
$\alpha_s(\frac{9}{4}m_D)=0.224$ \\ 3B & 0.290 &
$\alpha_s(\frac{9}{4}m_{\eta '})=0.291$ \\ 3B & 0.341 &
$\alpha_s(\frac{3}{2}m_\Phi ) =0.344$ \\ 3B & 0.356 &
$\alpha_s(\frac{3}{2}m_{\eta '})=0.356$ \\ 3B & 0.412 &
$\alpha_s(\frac{3}{2}m_\omega )=0.413$
\\ \hline  \end{tabular}

\vspace{0.5cm}

\caption{Local minima of the total vacuum energy $H_-(a)$
of the $N=3$ strings and comparison with strong couplings at energy levels
given by mesonic states.}

\end{table}

\begin{table}

\begin{tabular}{|l|c|l|} \hline

string & local minimum of $H_+(a)$ & strong coupling \\ \hline 3A
& 0.233 &$\alpha_s (\frac{2}{3}m_{\Lambda_b})=0.233$  \\ 3A &
0.282 & $\alpha_s(m_{\Lambda_c})=0.282$ \\
 3A & 0.508 &
$\alpha_s(m_{p,n})=0.508$
\\ 3A & 0.609 & $\alpha_s(\frac{2}{3}m_\Delta )=0.60$ \\ 3A &
0.628 & $\alpha_s(\frac{2}{3}m_\Sigma )=0.63$ \\ 3A & 0.70 &
$\alpha_s(\frac{2}{3}m_\Lambda )=0.68$ \\ 3B & 0.356 &
$\alpha_s(m_{N(1440)})=0.356$ \\ 3B & 0.367 &
$\alpha_2(m_{\Sigma^*})=0.366$ \\ 3B & 0.378 & $\alpha_s(m_\Xi
)=0.379$ \\ \hline

\end{tabular}

\vspace{0.5cm}

\caption{Local minima of the total vacuum energy $H_+(a)$ of the
$N=3$ strings and comparison with strong couplings
at energy levels given by baryonic states.}

\end{table}
As explained in section~9.1, the relevant energy scales for mesons
$M$ are given by $\frac{3}{2}m_M$ and $\frac{9}{4}m_M$, those for
baryons $B$ by $\frac{2}{3}m_B$ and $m_B$. Apparently, for
confined states the total vacuum energy $H_\pm (a)$ is the
relevant quantity to look at, rather than the single vacuum
energies $V(a)$ and $W(a)$.

Another interesting observation is that from the chaotic string
spectrum there is no straightforward evidence for supersymmetric
particles with masses in the region 100-1000 GeV. There are simply
no minima observed in the relevant energy region! Also, the
observed minima of the vacuum energy of the strings seem to
support grand unification scenarios based on the
non-supersymmetric beta functions, rather than the supersymmetric
ones. More on supersymmetry and possible grand unification
scenarios can be found in chapter 10.

In total, we found more than 30 zeros and minima that could be
identified with known standard model and gravitational interaction
strengths in a straightforward way. Moreover, more than 20 minima
could be identified with hadronic states. Could all this be a
random coincidence? It can't. Let us estimate the probability to
obtain all this by a pure random coincidence. In Tab.~1--3
we observe some 20 string couplings that coincide with
experimentally measured standard model coupling constants with a
precision of 3-4 digits. The joint probability for this being the
result of a pure random coincidence is of the order
$(10^{-3})^{20}=10^{-60}$. Here we still haven't taken into
account the hadronic minima of Tab.~5 and 6. Even if we
allow for different attributions of the minima to various possible
standard model interaction states, the joint probability of
obtaining randomly a joint coincidence of this order of magnitude
is still extremely small. In other words, a random coincidence can
be excluded. For this reason the chaotic string dynamics needs to
be integrated into future theories of particle physics in one way
or another, there is no way around it. The question of real
interest is whether chaotic strings are already the exact and
complete theory, or whether they are just the beginning (and
perhaps the perturbative approximation) of a more advanced theory.

\section*{12.6 Physical embedding}

We saw that the chaotic strings resemble the mass spectrum of
elementary particles in a correct way. That means, although a
complete picture certainly requires the field equations of QED,
QCD, weak interactions, Einstein's gravity, and possibly
superstring theory, the information on masses and coupling
strengths at a certain energy scale is correctly encoded by the
strings. The chaotic strings seem to contain the most important
information on what is going on at a certain temperature. In fact
they yield information on parameters that is not directly given by
the other theories, thus providing a nice and necessary amendment.

How can we understand the general role of these strings? Let us
become a bit philosophical and make a simple comparison. Suppose
you want to build a house. Before you can start, an architect has
to draw a plan of the house. The plan contains the most important
information about the house, but is certainly not the house
itself. Moreover, the plan is 2-dimensional, whereas the house is
3-dimensional. Also, the plan is drawn {\em before} the house is
being built. In our case, the plan are the chaotic strings and the
house is the universe. The strings are 1+1 dimensional, the
universe is 4 or 11-dimensional. The chaotic strings presumably
fix low-energy standard model parameters of the universe already
at an extremely early stage, long before standard model field
equations and a 4-dimensional space-time become relevant. The
different dimensionalities somewhat remind us of the holographic principle
\cite{susskind, maldacena,polyakov}, which relates degrees of freedom
of quantum field theories in different dimensions.

Keeping on being philosophical, we may even see some analogies
with biological systems. The DNA string encodes the most important
information about a living being, in fact already long before this
living being is being born. This information is encoded using
symbol sequences made up of 4 different symbols, which are
the chemical substances called Adenine, Cytosine, Guanine, and Thymine.
So for biological systems a kind of $N=4$ string is realized. The
DNA string is certainly not the living being itself, but it
contains the most important information about it. In that sense,
we may regard the chaotic string as a kind of `DNA' string of the
universe. It encodes the most important information on the
universe, and this may be relevant already long before ordinary
space-time is being created. What chemistry is for biology, is
information theory for physics.

Still one may ask how to concretely embed the chaotic string
dynamics into ordinary physics and possible theories of quantum
gravity. Several approaches are possible, all of which are
somewhat related.

The first and simplest approach is to regard the chaotic dynamics
as a new and {\em a priori} standard model-independent dynamics of
virtual momenta. We have shown in chapter 1 that it effectively
reduces to spatio-temporal Gaussian white noise if seen on a large
scale. This noise can then be coupled to the classical standard
model field equations and, in fact, can be used to generate the
noise fields of the stochastic quantization method. So the strings
are embedded as a tool for quantization. The new thing is that the
dynamics used for quantization is a deterministic chaotic one,
rather than a purely random one.

The second approach is to relate the chaotic dynamics to an
effective thermodynamic description of vacuum fluctuations allowed
by the uncertainty relation. Clearly ordinary statistical
mechanics is not valid for a description of vacuum fluctuations,
but the chaotic strings and their expectation values with respect
to the invariant measures of the coupled dynamics may yield the
correct tools for a statistical mechanics of vacuum fluctuations.
In fact, free ($a=0$) chaotic strings generate invariant densities
that can be regarded as generalized canonical distributions in the
formalism of non-extensive statistical mechanics, with an entropic
index $q$ given by either $q=-1$ or $q=3$. So the strings are
embedded as a generalization of statistical mechanics (more
details in chapter 6).

The third approach is to relate the chaotic strings to the
structure of space-time itself. There are 6 relevant chaotic
strings, which we labeled as $2A,2B,2A^-,2B^-,3A,3B$. Also, there
are 6 compactified dimensions necessary for the formulation of
superstring theory or M-theory. If we let the 6 chaotic strings
wind around (or even span up) the 6 compactified dimensions, then
they do not `disturb' our usual understanding of 4-dimensional
space-time physics. Rather, they yield a very relevant amendment.
Each coupling constant $a$ can then be regarded as a kind of
metric in the compactified space, and the analogue of the Einstein
field equations makes the observed standard model parameters
evolve to the minima of the effective potentials of the chaotic
strings. In this picture the chaotic strings are related to a
higher-dimensional extension of ordinary 4-dimensional space-time,
in fact to kind of `excited states' of ordinary 4-dimensional
space-time. The `ground state' (4-dimensional Minkowski space) is
represented by the (non-chaotic) $N=1$ strings, whereas the
chaotic strings with $N\geq 2$ span up higher states (higher
dimensions), similar to the energy levels of a quantum mechanical
harmonic oscillator with quantum number $N$ (more details on this
kind of approach in chapter 11).

Which of these different physical embeddings is the most relevant
one is not clear at the moment. Perhaps there is some truth in a
combination of all three of them.

\section*{12.7 Conclusion}

Instead of considering the standard model alone and putting in
about 25 free parameters by hand, in this book we have
postulated the existence of chaotic strings.
The chaotic string dynamics can be physically interpreted as a
1-dimensional strongly fluctuating dynamics of vacuum
fluctuations. It generates effective potentials which distinguish
the observed standard model couplings from arbitrary ones. The
dynamics may have already determined the standard model parameters
at a very early stage of the universe (in a pre-Planck scenario)
and it may still evolve today and stabilize the observed values of
the parameters.


There are 6 relevant chaotic string theories (Figs.~3, 4).
\begin{figure}
\epsfxsize=1. \hsize
\epsfig{file=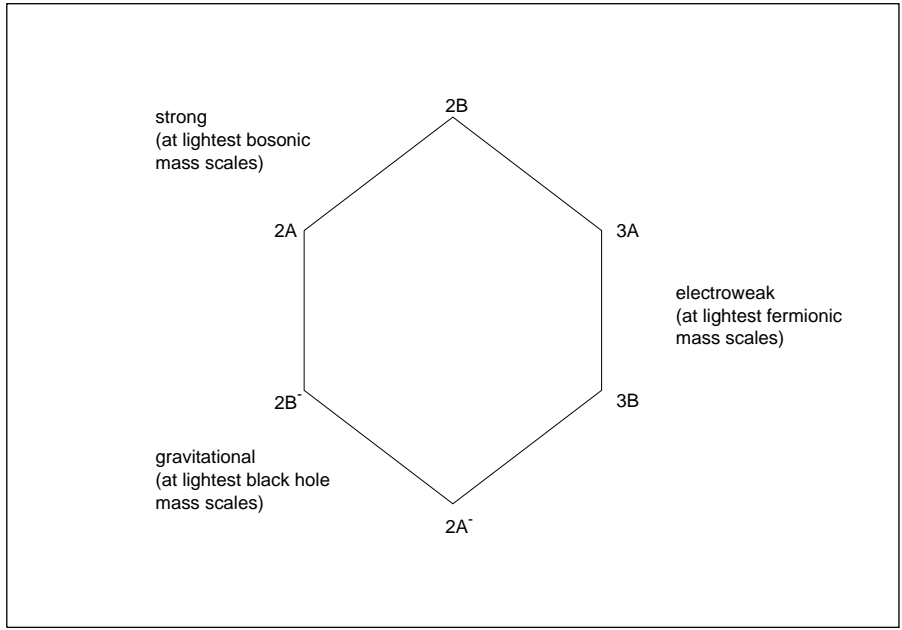}
\caption{The way in which the interaction energies $W(a)$ of the six chaotic string theories
fix the coupling strengths of the four interactions.}
\end{figure}
\begin{figure}
\epsfxsize=1. \hsize
\epsfig{file=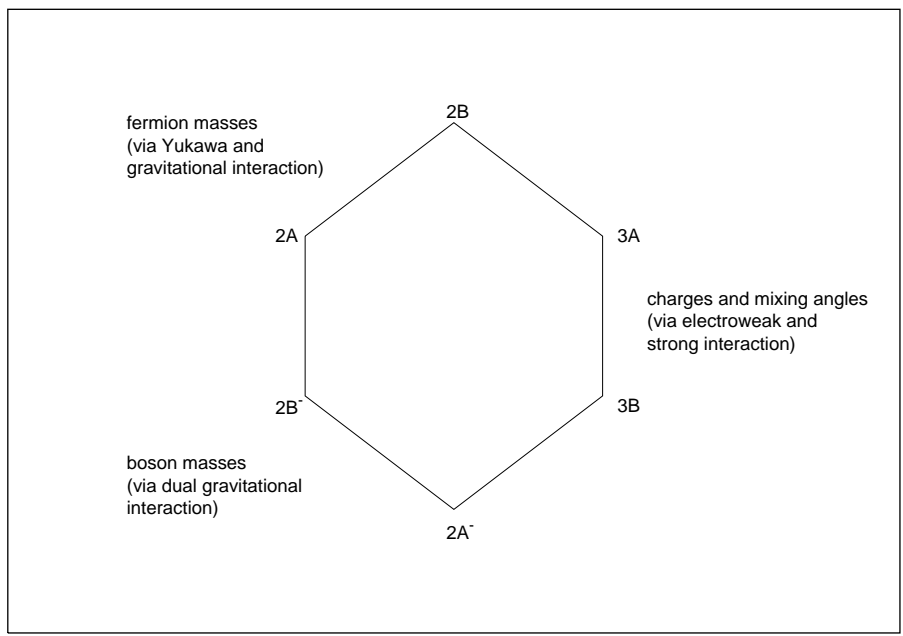}
\caption{The way in which the self energies $V(a)$ of the
six chaotic string theories fix the
charges, mass mixing angles and masses of the standard model particles.}
\end{figure}
Whereas for standard model fields, as well as for superstrings
after compactification, continuous gauge symmetries such as
$U(1)$, $SU(2)$ or $SU(3)$ are relevant, for the chaotic strings a
discrete $Z_2$ symmetry is relevant. In fact, the total theory may
be regarded as a $SU(3)\times SU(2)\times U(1)\times Z_2$ theory.
Whereas standard model fields or ordinary strings usually evolve
in a regular way, the chaotic strings obtained for $N>1$ evolve in
a deterministic chaotic (turbulent) way. They arise out of strongly
self-interacting 1-dimensional field theories and correspond to a
Bernoulli shift of information for vanishing spatial coupling $a$.
The constraint conditions on the vacuum energy (or the analogues
of the Einstein and scalar field equations) fix certain
equilibrium metrics in string space, which determine the strength
of the Laplacian coupling. We have provided extensive numerical
evidence that these equilibrium metrics reproduce the free
standard model parameters with very high precision (see
Tab.~4). Essentially coupling constants are fixed by the
interaction energy $W(a)$, and masses, mass mixing angles and
charges by the self energy $V(a)$. This is summarized in Fig.~3
and 4.

The simplest physical interpretation is to regard the chaotic
string dynamics as a dynamics of vacuum fluctuations, which is
present everywhere but which is unobservable due to the
uncertainty relation. Only expectations of the dynamics can be
measured, in terms of the fundamental constants of nature. The
strings can be related to a generalized statistical mechanics
description of vacuum fluctuations and may possibly wind around
the compactified space of superstring theory.

Generically, chaotic strings exhibit symmetry under the
replacement $V \to -V$ (or $T_N \to -T_N$), which can be formally
associated with a kind of supersymmetry transformation. However,
when introducing the evolution equations (\ref{sum101}) and
(\ref{sum102}) of the couplings one has to decide on the sign of
the constant $const$. This choice effectively breaks
the symmetry. Generally, with such a choice of sign the
expectation of the vacuum energy of the chaotic strings singles
out the physically relevant vacua, in the sense of stability.
Supersymmetric partners of ordinary particles, if they exist at
all, can be formally described by maxima rather than minima of the
effective potentials. But they are unstable with respect to the
fictitious time evolution, at least in our world. The instability
might indicate that supersymmetric partners, though formally there
to cancel divergences in the Feynman diagrams as well as unwanted
vacuum energy, may turn out to be unobservable in our world.

In any case, chaotic dynamics of the type studied in this book
seems to significantly enrich our understanding of standard model
parameters and of quantum fluctuations in general. Assuming that
on a very small scale quantum fluctuations are a deterministic
chaotic process rather than a pure random process the most
important free parameters of the standard model can be understood
with high precision. Embedding the chaotic strings into the
compactified space of a 10- or 11-dimensional theory we seem to be
looking at an extremely early stage of the universe where neither
matter nor radiation but information is the relevant concept.